\documentclass[lettersize,journal]{IEEEtran}
\usepackage{amsmath,amsfonts}
\usepackage{algorithmic}
\usepackage{array}
\usepackage[caption=false,font=normalsize,labelfont=sf,textfont=sf]{subfig}
\usepackage{textcomp}
\usepackage{stfloats}
\usepackage{url}
\usepackage{xcolor}
\usepackage{verbatim}
\usepackage{graphicx}
\def\BibTeX{{\rm B\kern-.05em{\sc i\kern-.025em b}\kern-.08em
    T\kern-.1667em\lower.7ex\hbox{E}\kern-.125emX}}
\usepackage{balance}

\begin{document}

\title{LiVeR: Lightweight Vehicle Detection and\\ Classification in Real-Time}

\author
{\IEEEauthorblockN
{Chandra Shekhar\IEEEauthorrefmark{1},
Jagnyashini Debadarshini\IEEEauthorrefmark{1},
Sudipta Saha\IEEEauthorrefmark{1}\\
\IEEEauthorblockA{\IEEEauthorrefmark{1}School of Electrical Sciences \\
Indian Institute of Technology Bhubaneswar, India\\ Email: \{cs13, jd12, sudipta\}@iitbbs.ac.in}}
}

\maketitle

\IEEEpeerreviewmaketitle

\begin{abstract}
Detection and classification of vehicles are very significant components in an \textit{Intelligent-Transportation System}. Existing solutions not only use heavy-weight and costly equipment, but also largely depend on constant cloud (Internet) connectivity, as well as adequate uninterrupted power-supply. Such dependencies make these solutions fundamentally impractical considering the possible adversities of outdoor environment as well as requirement of correlated wide-area operation. For practical use, apart from being technically sound and accurate, a solution has to be lightweight, cost-effective, easy-to-install, flexible as well as supporting efficient time-correlated coverage over large area. In this work we propose an IoT-assisted strategy to fulfil all these goals together. 
We adopt a top-down approach where we first introduce a lightweight framework for time-correlated low-cost wide-area measurement and then reuse the concept for developing the individual measurement units. Our extensive outdoor measurement studies and trace-based simulation on the empirical data show about 98\% accuracy in vehicle detection and upto 93\% of accuracy in classification of the vehicles over moderately busy urban roads.
\end{abstract}
\section{Introduction}\label{sec:introduction}

Efficient management of traffic is an extremely significant issue in the context of a smart-city or an intelligent-transportation system. Real-time understanding of the traffic through dynamic/on-the-fly classification of the vehicles is a crucial step towards that \cite{application}. Existing works to achieve the same are either based on (i) \textit{camera}, (ii) \textit{special-sensors}, or (iii) \textit{RF}. A major fraction of the works use still-image/video camera \cite{camera} and exploit computer-vision to accomplish the goal. These works need to employ a set of costly equipment over road-side and assume the availability of constant power supply to operate \cite{power}. They also use sophisticated ML/AI tools for image/video based detection of the vehicles and their classification which also require either presence of powerful computing facility on the spot or good network bandwidth for dynamic assistance from cloud servers \cite{its-cloud}.

Special sensors such as, loop-detectors \cite{loop-detector}, accelerometer \cite{accelerometer-ITS}, magnetometer \cite{magnetometer}, infrared \cite{infrared}, acoustic \cite{horn-ok-please} and several others have been used for vehicle detection and classification. However these works need special road-side installation support and substantial assistance from cloud to carry out the job. Different RF technologies including ZigBee \cite{RSSI-BASED-VTC}, WiFi \cite{VTC-MASS,witraffic}, BLE \cite{Bluetooth}, have been also used for the same purpose. Vehicles passing over the roads can create an obstructing effect over the RF signals. RF based works mainly exploit this fact for detection and classification of the vehicles.



For a vehicle detection and classification solution to be practically useful, it is anticipated to bear a set of salient features as listed below. (a) The solution should be expandable to cover a \textit{wide-area} and support easy \textit{coordination} among all the units. (b) \textit{Cost} involved to develop each unit should be low enough so that wide-area coverage is feasible. (c) It should be \textit{flexible} and can get easily installed in outdoor environments despite of the possible hostilities and \textit{should not depend on any special infrastructure}. (d) The processing should be simple enough so that use of cloud is minimized. (e) It should be able to operate using battery-power as and when needed. (f) All the units should be able to collect /exchange \textit{time-correlated data in real-time} with each other for smooth system-wide coordination. Detailed study of the existing works (see next section) reveals that, while each of them focus on certain specific  issues, in general they fail to satisfy all the above mentioned requirements together.


In this work we propose an IoT-assisted low-power RF based lightweight solution that addresses most of the above mentioned conditions. Our work mainly exploits the characteristics of the disruptions in the low-power RF communication links in IoT caused by the vehicles \cite{comsnets-mitigation, volume-measurement}. In contrast to the existing works that study RF \cite{sensys,Bajwa-pavement} through a very controlled and localized setting, in the current work we demonstrate design and development of a complete stand-alone system to solve the whole problem satisfying most of the prime requirements.


Fundamentally, there are two challenges in solving the problem using low-power RF. (i) Degradation of the quality of links is used as the foundation of \textit{detection and classification}. Although it goes in favor with the low-power RF communication, at the same time, it becomes challenging to use the same mechanism to develop a complete protocol with support for end-to-end coordination. (ii) A sufficient number of packets require to collide with a vehicle and the quality of their reception is to be analyzed on-the-fly for such a strategy to work. However, this is hard without any controlled-setting/specialized arrangements since due to the mobility of the vehicles, the time for which a single vehicle can obstruct a communication link is very low.

Under traditional \textit{Asynchronous-Transmission} (AT) based strategies, its quite hard to address these issues while at the same time keeping the strategy lightweight, real-time as well as support end-to-end coordination over wide-area. In this work we propose a \textit{Synchronous-Transmission} (ST) based framework to solve the problem. ST has gained quite a good popularity with several recent studies \cite{LWB,chaos} in the context of IoT/WSN because of its ability to achieve quite high reliability under low-latency and low-radio-on time. 
In a nutshell, in contrast to the existing works, here we adopt a top-down approach where we first focus towards design and development of an ST based efficient framework for time-correlated real-time measurement over wide area and next address the issue of cost-effective and efficient measurement in each unit.



The main contributions from the work can be summarized as follows.

\begin{itemize}
    
    \item An ST based framework has been proposed for time-correlated detection and classification of vehicles in real-time over wide-area.
    
    \item Lightweight ML-models have been used in low-power low-cost off-the-shelf IoT-edge-devices for on-the-fly detection as well as classification of the vehicles into three classes, namely, \textit{small}, \textit{medium} and \textit{large} (based on size) without any assistance from cloud.
    
    \item The proposed system is implemented in Contiki OS for TelosB devices and extensively studied through network emulation platform as well as outdoor experiments over multiple different places over a city.
    
    \item Our trained model exhibits a detection accuracy of 98.8\% with quite descent classification accuracy of 91.3\%, 92.3\% and 93.8\% for small, medium and large size vehicles, respectively.
    
    
\end{itemize}

The rest of the paper is organized as follows. Section \ref{sec:relatedworks} provides a summary of the related works. Section \ref{sec:background} reports a brief description of the background. Section \ref{sec:design} provides a detailed description of the design of the proposed framework. Metrics/features used in the proposed strategy have been discussed in Section \ref{sec:metric}. 
Subsequently, Section \ref{sec:measurement} and \ref{sec:sensitivity-detection} describe the efficacy and efficiency of the proposed strategy for detection of the vehicles. Section \ref{sec:vehicle} details the design of the proposed vehicle classification strategy and Section \ref{sec:evaluation} provides the evaluation study of the same in outdoor settings. Finally, Section \ref{sec:widearea} provides an network emulation based study of the proposed strategy and its ability of wide-area time-correlated operation.

\section{Related Works}\label{sec:relatedworks}

Here we provide a comprehensive summary of the existing works on vehicle detection and classification with respect to the expected features as discussed in the Section \ref{sec:introduction}.



\textbf{Cost-effectiveness:} The first step towards a cost-effective deployment of a complete system covering a large area, is to have a cost-optimized implementation of a single unit. Video/images \cite{camera,camera-CNN,camera-couting} based solutions use costly hardware equipment for recording of the vehicles. Sensor assisted solutions also mostly need to depend on installation of costly infrastructure \cite{survey-sensors} for appropriate functioning of the sensors. In principle, RF based solutions do not need any such costly equipment \cite{Bluetooth,RSSI-based,RSSI-BASED-VTC}. However, the way RF has been used so far, it also involves high cost per unit for the necessary setup for controlled measurement of physical layer parameters (e.g., \textit{signal strength}) \cite{witraffic,VTC-MASS,VTC-2020}. In contrast, in this work we use low-power RF communication between off-the-shelf IoT-devices and in-device recording as well as lightweight analysis of the data which by default optimizes the cost per unit.

\textbf{Flexibility/Infrastructure:} Vehicle detection and classification task is mostly supposed to be carried out in an outdoor environment. Unlike friendly indoor environment, an outdoor is considered to be quite hostile and dynamic. Thus, flexibility and easy re-installation support for a solution is very important. Camera-based solutions naturally need special installation of the costly-camera and their proper maintenance. Sensor assisted solutions also need specialized installation of the sensors. For instance, the intrusive sensors such as acoustic-sensor \cite{horn-ok-please}, light-sensor \cite{optical}, magnetometer \cite{magnetometer,magnet-2}, loop detector \cite{loop-detector}, piezoelectric sensor \cite{Piezo-Electric-Sensor}, vibration \cite{vibration,vibration-2} etc., first need to get appropriately installed on the pavement surfaces. High installation costs, traffic disruption during installation, maintenance, etc., are some of the disadvantage of such approach. In contrast, in the current work we use off-the-shelf battery-operated small-size IoT-devices to serve the purpose. The IoT-devices can be installed at any locations facing the road, e.g., road-side tree/lamp-posts etc. It does not need any special structure and hence supports flexibility and easy re-installation quite well.

\textbf{Energy-requirement/Power-supply:} In general all the existing works assumes availability of adequate and uninterrupted source of power (e.g., wall-power) for carrying out the job. In contrast, our design exploits low-power IoT-systems and hence, largely relaxes this requirement. In particular, the framework we propose here can sustain its operation for a substantial amount of time over battery power. Moreover, the system can be appropriately tuned to run over battery power for longer time also.

\textbf{Cloud Assistance/Computation-cost:} All the Camera based solutions \cite{camera-CNN,camera-couting,camera,infrared} need to run complex algorithms in the back-end for processing of the images/video data. To carry out such processing, constant cloud connectivity is a must. Frequent communication with cloud with large data creates a serious bottleneck in mass application of such solution over a large area. As a common trend in the current design policies in general the interaction with the cloud and the edge systems is supposed to be minimal \cite{cloud-load}. In the IoT-assisted approach proposed in this work, we keep the process very simple so that it can be performed in low-power small-size IoT-devices itself without any explicit requirement of cloud connectivity.

\textbf{Wide-area Synchronized measurement:} Most of the existing works mainly discuss about the detection and classification of the vehicles at a single place \cite{Bluetooth,VTC-MASS}. However, in the context of a smart-city/intelligent-transportation, the solution needs to be not only replicated at multiple locations, but also the measurements carried out at these different locations need to be time-correlated to support fruitful on-the-fly system-wide analysis. Our proposed IoT-based framework, apart from being a cost-effective and easy-to-use solution, takes care of these vital issues which is largely missing in the existing works. Intrinsically, our proposed framework does not depend on the conventional 3G/4G/LTE network. Rather it is designed as an ST based \textit{independent} IoT-system connecting all its units carrying out measurements at different locations, thus inherently supporting the time-correlated measurements.


\textbf{System-wide analysis:} The primary goal of wide-area detection and classification is to enable the traffic management system to take appropriate decision dynamically. 
In the existing works, the only option is to upload the data collected from each unit to a cloud server, analyze them centrally and take decision. However, conveying data to the cloud at every time epoch from every measurement unit installed over a large area would naturally result in huge bandwidth consumption, causing network-congestion and in turn heavily degrading the real-time flavor of a solution. None of the existing works pay attention to this serious concern \cite{Bluetooth,VTC-MASS,VTC-2020}. The proposed IoT-assisted solution in the current work, in contrast, by the virtue of the existing rich IoT-protocol-base \cite{glossy,chaos}, can carry out in-network data processing to support on-the-fly end-to-end collaborative decision making in real-time without any active help from cloud ensuring much faster traffic management and control.




\begin{table*}
\begin{center}
\includegraphics[angle=0,width=1.05\textwidth, height = 6.5 cm]{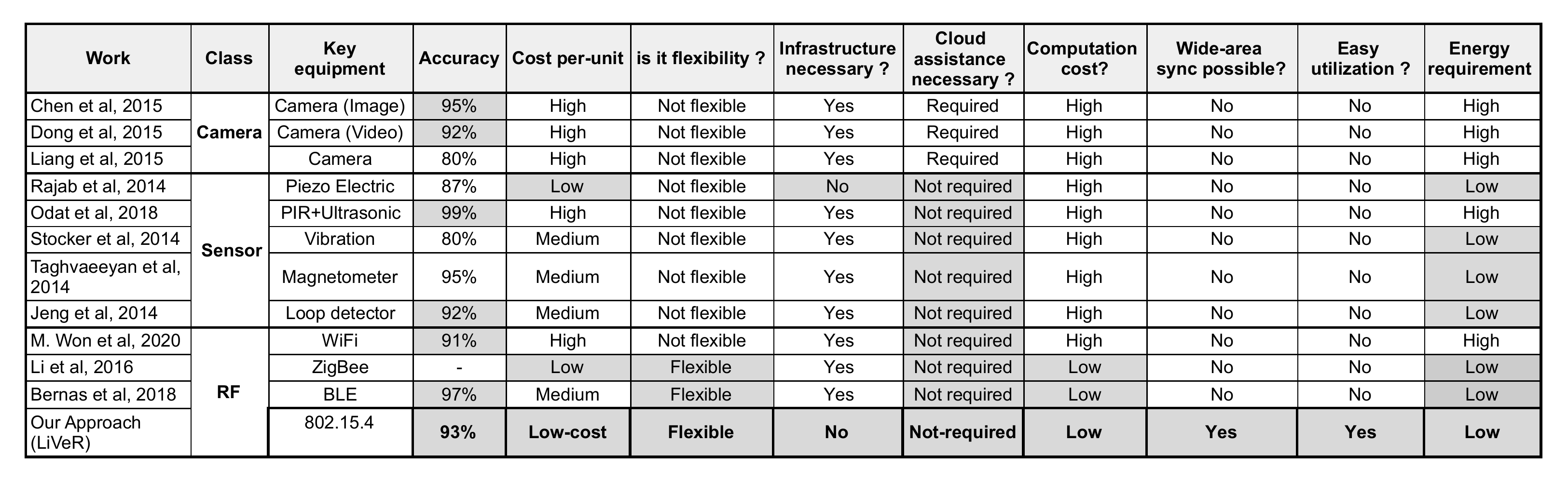}
\end{center}
\caption{Summary of the existing works on vehicle detection and classification.}
\label{table:Review}

\end{table*}



Table \ref{table:Review} shows the summary of the review of the existing works. Each row in the table talks about the availability of the necessary features in an individual work. The specific issues that are favorable towards its practical deployment is marked with gray background. In a nutshell, it is found that although each of the existing solutions, has certain positive side(s), almost none of them addresses all the desired qualities together.

\section{Background}\label{sec:background}

Communication hindrance in a low-power IoT-systems caused by the large vehicles passing over the roads have been reported by several works so far \cite{sensys,comsnets-mitigation}. Such adversity motivates us to use a low-power communication protocol as a cost-effective and lightweight ``sensor" for vehicles and their classification. However, at the same, possibility of severe disturbances caused by the vehicles fetches significant challenge in developing a full system solely using a low-power technology. Essentially, it points towards devising a way where first, we have a robust low-power communication protocol and next, the vehicular adversity is allowed to cause deterioration to the protocol in a very controlled way.

Under traditional AT based communication in a WSN/IoT system, to carry out sustained low-power operation the devices require to do duty-cycling (\textit{periodic ON and OFF}) of their RF-transceiver unit since it consumes the maximum power in the device. For a network-wide end-to-end communication, the nodes gradually get tied up with each others' duty-cycle in a systematic fashion. For instance in the well-known data-collection protocol CTP \cite{CTP} a source node syncs its duty-cycle with its immediate parent, i.e., the node that is assigned the responsibility to forward its data towards the sink node. Although WSN/IoT systems operate under a wireless broadcast medium, such pair-wise tying up highly limits the connectivity among the nodes which makes the system more prone to temporal link disturbances, e.g., disturbances caused by the vehicular obstacles in an urban environment.

In contrast, ST-based protocols are organized in an absolutely different way. The devices, under ST, try their best to avoid collision among packets with the help of physical layer phenomena called \textit{Constructive-Interference} (CI) or \textit{Capture-Effect} (CE)\cite{capture-effect}. However, for this to happen the protocol mainly requires the data packets having identical content to be transmitted by every node. Therefore, usually flooding of a data unit is considered to be the basic unit of communication under ST. 

\textbf{Glossy:} The pioneering protocol Glossy \cite{glossy} shows how flooding can be achieved in a very compact form using ST. The process starts with a single node broadcasting the data packet to be disseminated over the system. It is received at the same time by all the first-hop neighbors of the initiator. Through appropriate time-adjustment, the protocol arranges the re-transmissions of the received packet from every first-hop node in a way so that instead of colliding they results in CI/CE and subsequently second-hop neighbors of the initiator successfully decode the packet. The process cascades this way until the full-network is covered. To ensure complete network coverage, every node repeats the transmission of a packet a pre-decided number times (referred to as NTX) with each transmission followed by a one reception.


An ST based flooding, thus, successfully exploits the strength of the redundant transmissions of packets from multiple different source nodes in the best possible way. This makes it inherently much more resilient to the degradation of the links caused by vehicular obstructions compared to the AT based strategies. In the next section we describe how we use ST-based flooding to build a framework to serve our purpose. 

\section{Design of the Framework}\label{sec:design}

The design of the framework is supposed to address two conflicting goals. \textit{Firstly}, we try to exploit the characteristics of the degradation of low-power communication links. \textit{Secondly}, we also aim to build a lightweight low-cost framework based on the same low-power communication technology. For the first goal, we need to make a certain number of packets collide with the vehicles \textit{in a controlled way} as well as analyze their reception status on-the-fly in the devices itself which is fundamentally impossible to arrange without the accomplishment of the second goal.

We use an ST-based \textit{top-down} approach to meet both the goals together. Unlike other existing works, we first think about design of a platform which can support a time-correlated operation over an wide-area. The ST-based fast flooding protocol Glossy is used to form the said platform. Next, inside this robust framework, in a controlled way, we inscribe an independent small module that allows itself to degrade as per the passage of the vehicles for a defined small period. In particular, for this purpose we use an independent instance of Glossy itself in a \textit{sensitive-mode} by sufficiently \textit{lowering the transmission power level}. The whole process repeats periodically so that the system keeps on running safely while producing enough scope for carrying out the necessary measurements.

\begin{figure}[htbp]
\begin{center}
\includegraphics[angle=0,width=0.5\textwidth]{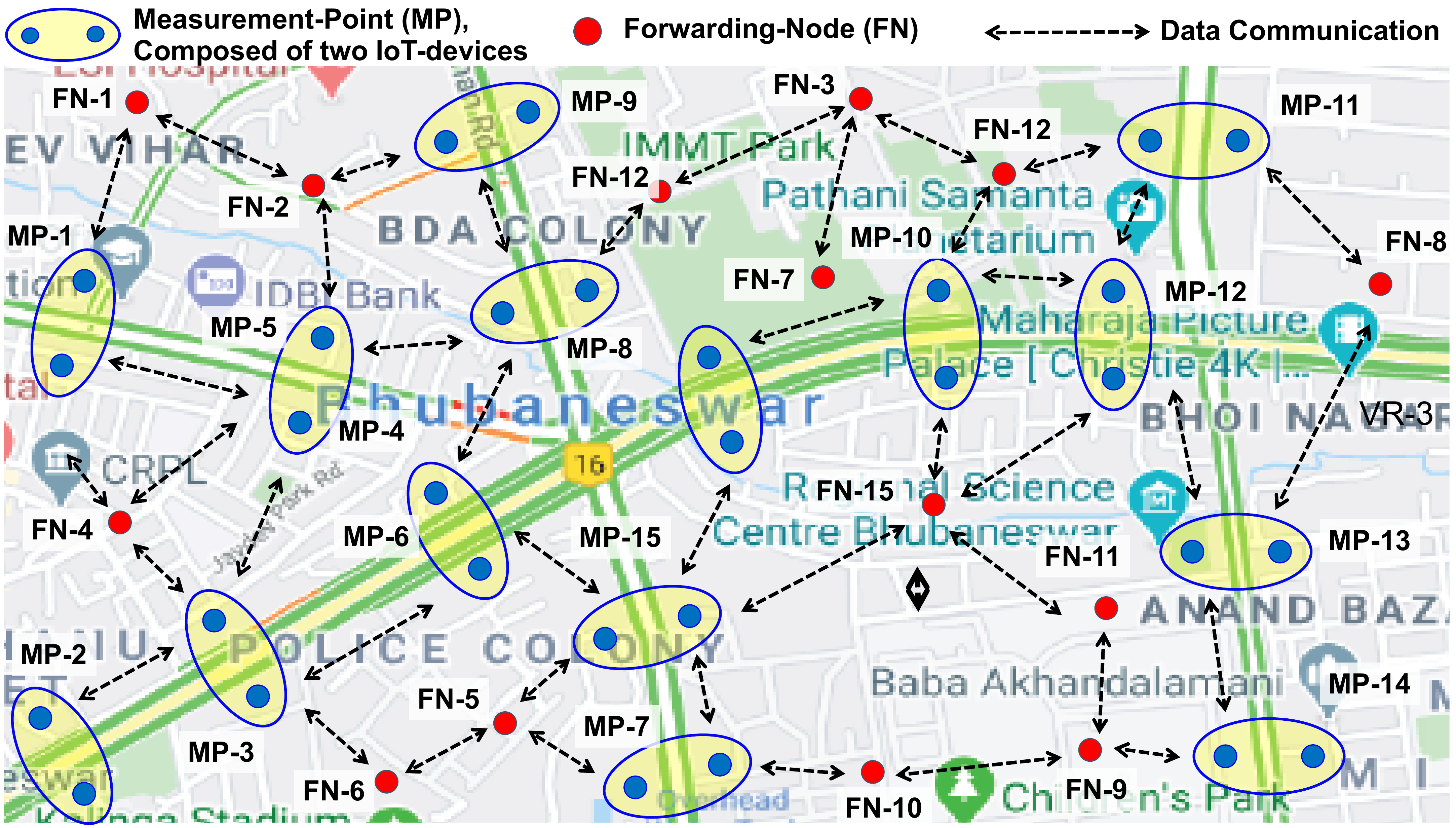}
\end{center}
\caption{Possible deployment of the Measurement-Points (MP) and Forwarding-Nodes (FN) over an area in a city.} \label{fig:ground}
\end{figure}

\begin{figure}[htbp]
\begin{center}
\includegraphics[angle=0,width=0.5\textwidth]{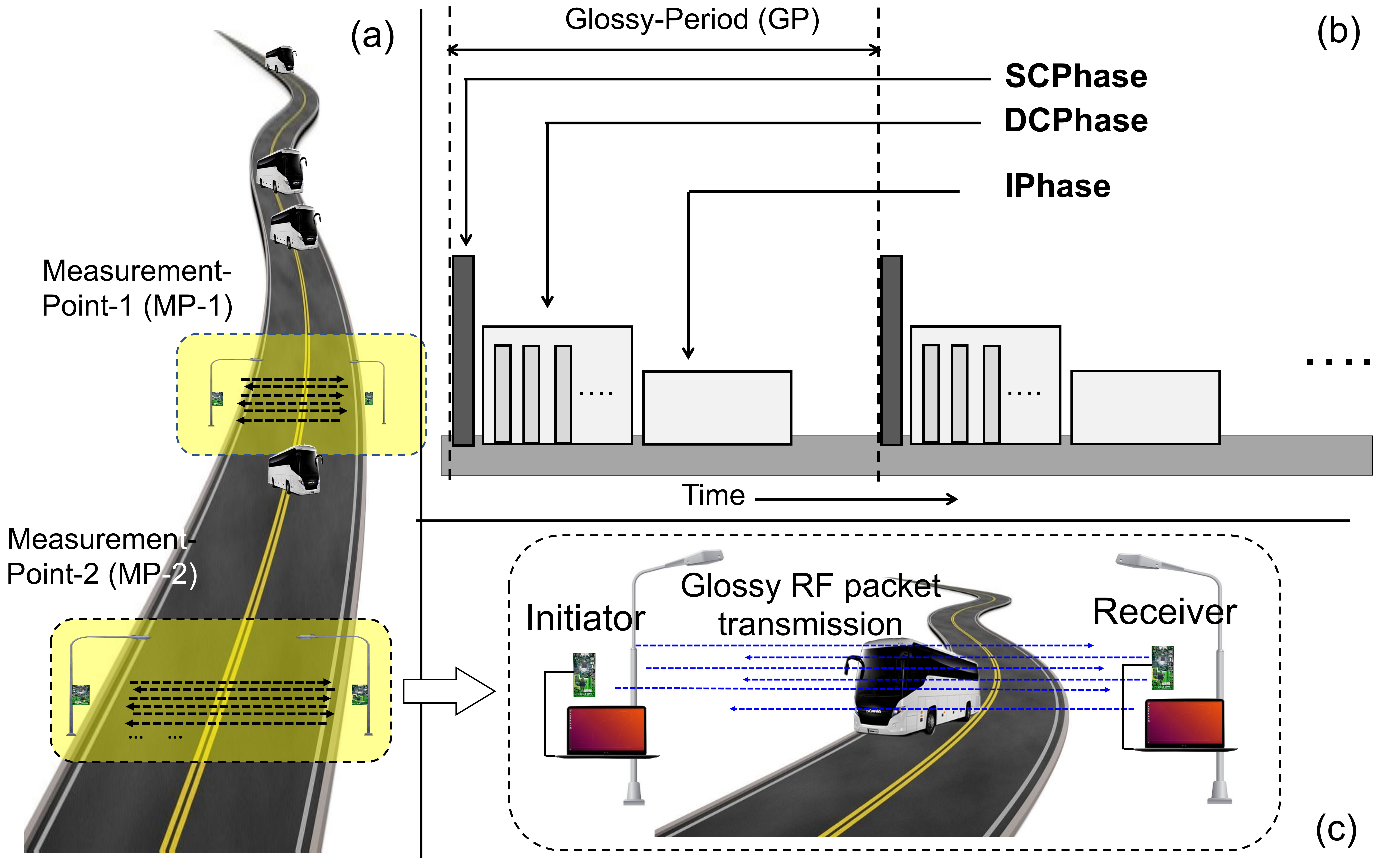}
\end{center}
\caption{ Part (a) (left side) of shows a road with two MPs (MP1 and MP2). Part (b) shows the time-diagram of the execution of the proposed three-phase framework. Part (c) shows the operation of an MP minutely, where the transmission of packets between the initiator and the receiver (shown by arrows) get partially or fully obstructed by the vehicles.} \label{fig:experiment}
\end{figure}

Figure \ref{fig:ground} shows a schematic of the deployment of the off-the-shelf IoT-devices over a possible target area. The devices can be deployed in either of the two forms - (a) as a \textit{Forwarding-Node} (denoted by FN-$i$ in the figure) or (b) as a part of a specially designed \textit{Measurement-Point} (MP) (denoted by MP-$i$ in the figure) as described later. An MP is composed of \textit{two low-power WSN/IoT-edge devices} installed at the two opposite sides of a road. Figure \ref{fig:experiment}(c) magnifies a single MP while Figure \ref{fig:experiment}(a) shows placement of two MPs over a road. The two devices in an MP are referred to as the \textit{initiator} and the \textit{receiver}. All the MPs and FN all together form a connected network/ IoT-system.

The overall process runs as a periodic repetition of the  \textit{three phases}: (i) In the \textit{first-step}, Glossy runs over the full system for time-synchronization and sharing of necessary control information. The robustness of Glossy ensures that the platform keeps on running despite random failures of the links at various location of the network due to vehicular obstructions. This phase is referred to as a \textit{Synchronization and Control Phase} \textit{(SCPhase)}. (ii) Subsequently, the process enter to the \textit{second-step}, where only the devices within an MP executes Glossy in a sensitive low-power mode. This is referred to as \textit{Detection and Classification Phase} \textit{(DCPhase)}. The pattern of the degradation is captured in the devices itself and is analyzed on-the-fly through a light-weight ML strategy to detect the presence of the vehicles and their classification. (iii) Finally, there is an optional \textit{third-step} referred to as \textit{Interaction-Phase} \textit{(IPhase)} where the MPs interact with each other and derive/share useful information for system-wide coordination and decision making.

\textbf{Glossy-Period (GP) \& Glossy-Duration (GD):} The three-phase process described above runs in a \textit{periodic fashion} with a defined period and defined duration for each of the three phases as per requirement. Figure \ref{fig:experiment}(b) pictorially depicts this arrangement in the form of a time-diagram. GP is the period set for the whole process. GD is part of the period where the actual data communication takes place. By the virtue of ST based mechanism Glossy inherently executes very fast. For instance, to cover a 100-node ZigBee network spread over about 5 hops, Glossy takes about 10 - 20 ms to achieve a full network-wide one-to-all dissemination. Thus, for SCPhase, GD is decided based on the size of the full network (we refer this as GD$_\text{s}$) while in DCPhase it is set based on the controlled time we would like the RF packets to interact with vehicles (referred as GD$_\text{d}$). GP is set only for the SCPhase. 

\section{Features/Metrics}
\label{sec:metric}

The whole concept behind the work is centered around controlled measurement of the characteristics of the degradation of the performance of Glossy in the DCPhase due to interaction with the passage of the vehicles. To serve the purpose, we intentionally make the protocol sensitive to vehicular disturbances by reducing the transmission power level appropriately so that it works normally when there is no traffic but degrades gracefully with the presence of traffic.


\textit{Physical-layer} (PHY) parameters such as RSSI, and LQI have been used in several prior works \cite{comsnets-lqi-prr,comsnets-mitigation,RSSI-BASED-VTC,RSSI-based} to characterize the changes in the RF communication during the passage of vehicles. However, in order to realize a lightweight and easier analysis of the measurement data in resource-constrained IoT-devices in this work we go for capturing the \textit{Data Link Layer} (DLL) parameters instead of the raw PHY parameters. Although disturbances in the DLL parameters is fundamentally an artifact of the same in the PHY parameters, moving to a higher layer in the network stack provides a natural filtering of the effects which in turn makes the process of analysis simpler and quicker. The DLL performance metrics which we use as a set of features in this study are detailed next.


\textbf{Timeout-Count (TC):} A Glossy initiator automatically makes multiple attempts to send a packet to the receiver in case it does not get any response back from it. TC refers to the total number of times the initiator tries. TC is ideally 1 without presence of any vehicle and increases depending on the degree of the obstruction. 


\textbf{Latency (LT):} It refers to the time taken by a receiver node (non-initiator) to successfully receive the data transmitted by the initiator. Note that reception of at least one packet out of NTX is enough for the receiver. In DCPhase, in absence of any vehicle LT is, thus, supposed to be very minimal. However, due to the failures in reception of the packets by the receiver due to vehicular obstruction, LT goes higher.


\textbf{Reception-Count (RXCT):} In Glossy, an initiator sends NTX packets. Hence, in an off-road setting, receiver is supposed to receive almost all the NTX packets. However, due to the vehicular obstruction, either some of the packets may get lost or all the transmissions might not get completed by the time GD$_\text{d}$. Thus, the number of packets received in the receiver side gets lesser than NTX due to vehicular obstructions which we also use as a feature.


\textbf{Radio-On time (RO):} RO is the time taken to complete all of the NTX number of packet transmissions by a node. In Glossy every node (except initiator) transmits only after a reception. If a packet gets lost it may be re-transmitted stretching the completion of the communication process within GD$_\text{d}$ or the process itself may fail to convey all NTX packets because of limited GD$_\text{d}$. To capture all sorts of such variations caused by vehicular obstruction we include RO also as a feature. 


While we mainly focus on the DLL metrics, the two primary PHY metrics RSSI and LQI are also logged for analysis. By the virtue of our proposed framework we successfully capture the RSSI and LQI in a discrete form (pair of values for each packet). We calculate the average and standard deviation of these RSSI and LQI values over all the packets during DCPhase in the devices. We refer to these as \textbf{RSSI-avg}, \textbf{RSSI-stddev}, \textbf{LQI-avg} and \textbf{LQI-stddev}.



\begin{figure*}[htbp]
\begin{center}
\includegraphics[width=\textwidth]{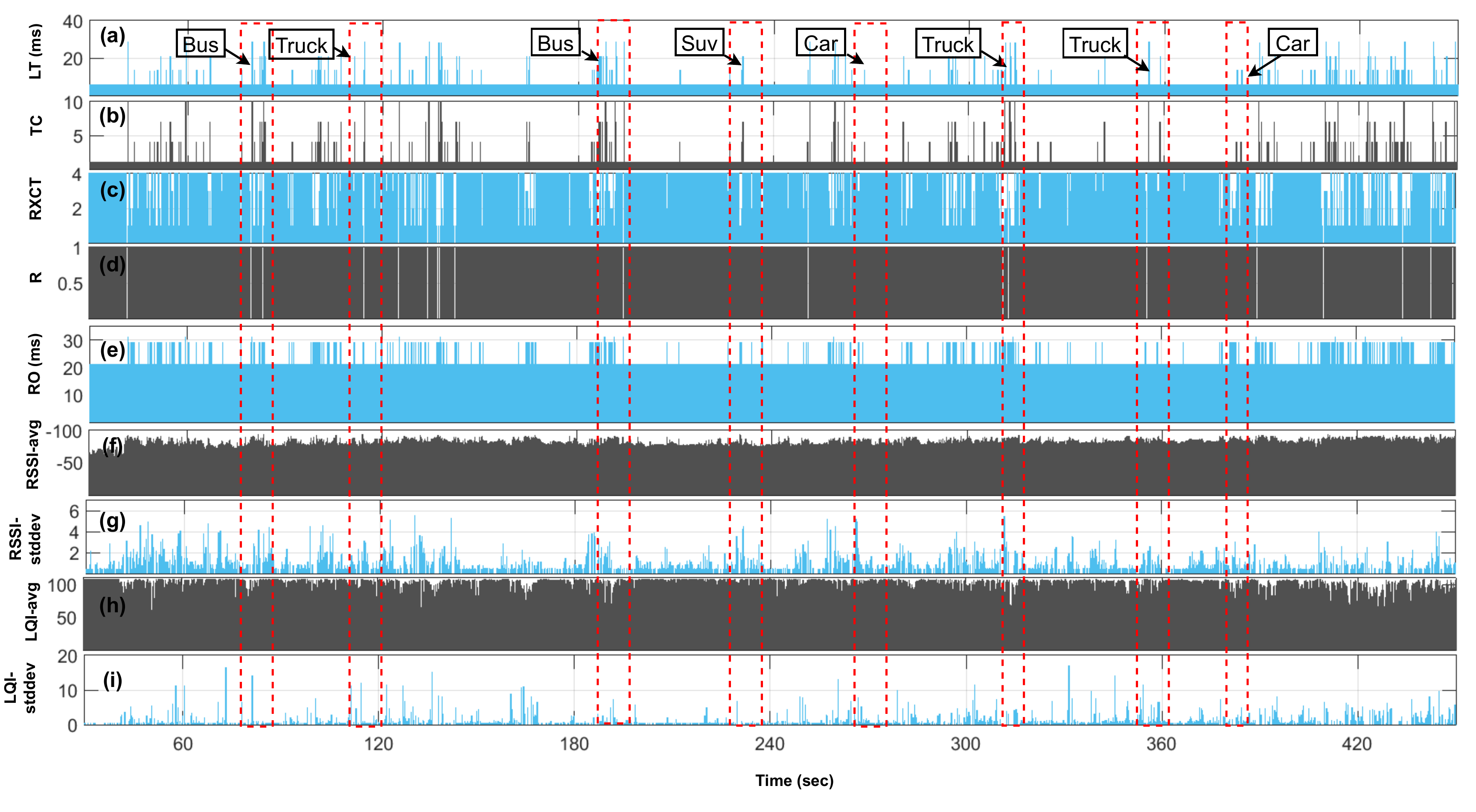}
\end{center}
\caption{Time-series representation of the metrics Latency (LT), Timeout-Count (TC), Reception-Count (RXCT), Reliability (R), Radio-on time (RO), RSSI-avg, RSSI-stddev, LQI-avg, and LQI-stddev.}
\vspace{-0.5cm}
\label{fig:timeseries}
\end{figure*}

\begin{figure}[htbp]
\begin{center}
\includegraphics[width=0.5\textwidth]{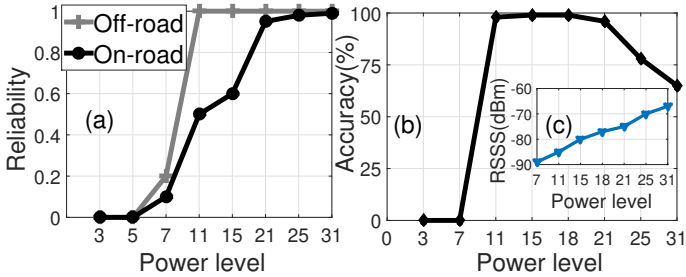}
\end{center}
\caption{Part (a) shows the reliability of Glossy at different power level in off-road and on-road setting. Part (b) shows the accuracy in detection of the passage of the vehicles using an SVM classifier trained with data collected at different transmission power levels.}
\vspace{-0.5cm}
\label{fig:pl-reliability-accuracy}
\end{figure}

\section{Measurement Study} \label{sec:measurement}

The proposed framework is implemented in Contiki OS for TelosB devices. We carry out an extensive outdoor link measurement study with a single MP composed of a of TelosB devices installed at the two opposite sides of a road at a distance of 12 meters from each other (as shown in Figure \ref{fig:experiment}(c)). The experiment is repeated at several places over the Bhubaneswar city. In our experimental setup, both the initiator and the receiver are connected with laptop for logging data (see Figure \ref{fig:experiment}(c)). Video of the whole measurement process is recorded as a \textit{Ground-Truth} (GT).   

Figure \ref{fig:timeseries} plots all the four DLL metrics and two PHY metrics over a seven-minute of measurement data. For an overall understanding we also measure the \textbf{Reliability (R)} of the Glossy instance in the DCPhase. Reliability of a Glossy instance is considered to be 100\% in case at least one of the NTX transmission attempts from the initiator reaches the receiver. Thus, value of R can be mostly 1. However, it may also go to 0 when the adversity is so serious that none of the NTX packets could reach the receiver. Figure \ref{fig:timeseries}(d) shows the plot of R.

We annotate the data based on GT and find that the readings of all the metrics successfully capture the disturbances caused by every passage of the vehicles through the MP in a very correlated way. Several such cases are shown using dashed rectangles in Figure \ref{fig:timeseries}. While, LT and TC both are found to be exhibiting a sharp and wide change over occurrences of almost all types of vehicles, RO is found to be showing mostly a binary pattern with a \textit{high} value indicating the presence of the vehicles and a \textit{low} value indicating absence. RXCT behaves in an exactly opposite way. When there is no vehicle, all the NTX transmissions and receptions complete fast and naturally RXCT goes high. Whereas, it goes low with the presence of the vehicles. 

It can be observed that although RSSI-avg hardly shows any correlation with the disturbances, RSSI-stddev shows a good match and also mostly correlates with TC. The same nature is visible in LQI-stddev. However, from visual inspection of the difference between the patterns produced by PHY metrics and DLL metrics w.r.t GT it can be understood that PHY metrics change in a more continuous form while DLL metrics appropriately discretizes the measurements which ultimately becomes very instrumental in simplifying the process of classification in resource constrained devices. 




\section{Vehicle Detection}
\label{sec:sensitivity-detection}

In the measurement experiments reported in the previous section, to make DCPhase sensitive and SCPhase resilient to the obstruction created by the passage of the vehicles, we set the \textit{Transmission-Power-Levels} (TXP) in these two phases as 15 and 31. These values are reported with respect to the power-levels of CC2420 (the 802.15.4 compatible radio used in TelosB). It allows 32 different power-levels starting from 0 to 31. Inset in Figure \ref{fig:pl-reliability-accuracy}(b) shows the received power in dBm at the receiver of an MP when a packet is transmitted at some of these power levels from the transmitter in our experimental setting.

The detection efficiency of the proposed framework strongly depends on the sensitivity of DCPhase w.r.t. the passage of the vehicles. Intuitively, a low-strength signal gets more affected due to the vehicular obstructions, while a higher strength signal acquires the ability to tolerate the disturbances. Thus, although low signal strength goes in our favor, setting too low power-level in DCPhase may have undesired effects due to either complete abolition of the packet or result in inadequate residue strength which may not get detected by the receiver. To get a clear understanding, we repeat the measurement study under different TXP in both \textit{off-road} (without traffic) and \textit{on-road} (with traffic, over a moderately busy road) setting. All the metrics are measured over 5000 iterations with transmission power level starting from 3 (-98 dBm at the receiver) upto 31 (highest, -66 dBm at the receiver as shown in Figure \ref{fig:pl-reliability-accuracy}(c)). 

Figure \ref{fig:pl-reliability-accuracy}(a) shows the plot of R in a general Glossy protocol. In off-road setting, a drastic drop is visible in R below TXP 11. Whereas above TXP 25, reliability is found to be staying above 99\% even in on-road setting. Based on this study, (i) we \textit{select TXP above 25 for SCPhase} to keep the system running despite of the vehicular obstructions, while (ii) \textit{set TXP between 11 and 25 for DCPhase} to induce controlled sensitivity.

\begin{figure}[htbp]
\begin{center}
\includegraphics[width=0.5\textwidth]{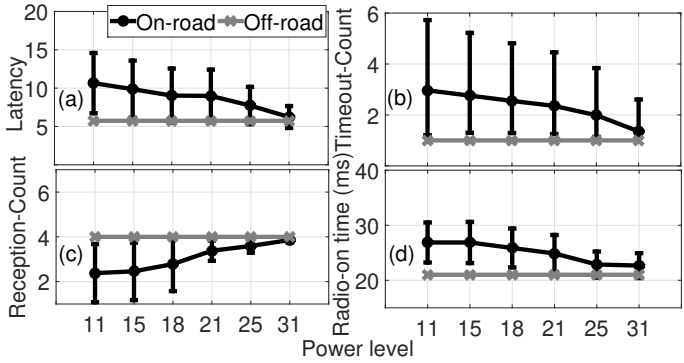}
\end{center}
\caption{Effect of TXP on the metrics Latency, Timeout-Count, Reception-count and Radio-On time measured in DCPhase.} \label{fig:motivation}
\end{figure}

We carry out extensive outdoor experiments with the same setup at a number of different power levels between 11 and 25 and measure all the metrics. Figure \ref{fig:motivation} shows the plots average LT, TC, RO, and RXCT and their standard deviation over all these experiments in both on-road and off-road. As visible from the results, at TXP above 11 all the metrics are quite stable in off-road setting and do not show any variation. However, all of them are found to be widely varying under on-road setting (as visible through the standard deviation/error-bars in Figure \ref{fig:motivation}). In general the lesser the TXP the wider the variation.

It can be inferred from the study above that any TXP above 11 (and below 25) is quite good for detection purpose. However, to automate the detection process based on the on-the-fly measurements of the metrics, we use a lightweight ML strategy known as SVM. The 80\% of the collected data are used for training an SVM based binary classifier. The rest of the data are used for recalling/testing. The process is repeated for each of the four metrics. Figure \ref{fig:pl-reliability-accuracy}(b) shows the accuracy (w.r.t. GT) 
results while recalling at different TXP using the metric LT. The accuracy is found to be quite high in the range of TXP 11 to 21. For TXP higher than 21, medium and lower size vehicles failed to make enough impact on the protocol. Similar behavior is observed for the other three metrics too.

\section{Vehicle Classification} \label{sec:vehicle}

With reasonable success in detection, we move for the next step, i.e., classification. Basic measurements studies as depicted in Figure \ref{fig:timeseries} quite clearly shows that the DCPhase gets affected differently by different type of vehicles. For instance, passage of a \textit{truck} results in more rise in LT, TC and more drop in RXCT than a \textit{car}. In general, the disturbances in almost all the metrics varies with the size/volume of the vehicle with which the DCPhase collides. In this section we study how well such differences can be captured and used for vehicle classification through the proposed lightweight framework. 

We define three classes of vehicles based on their sizes: (a)\textit{Type-S}: These are small-sized vehicles, e.g., van, auto, etc. (b) \textit{Type-M}: These are medium-sized vehicles, e.g., SUV, big-size car etc., and (c) \textit{Type-L}: These are large-sized vehicles, e.g., truck, bus, etc. Notation \textit{Type-N} is used to indicate \textit{no-vehicle present}. 

\begin{figure}[htbp]
\begin{center}
\includegraphics[width=0.5\textwidth]{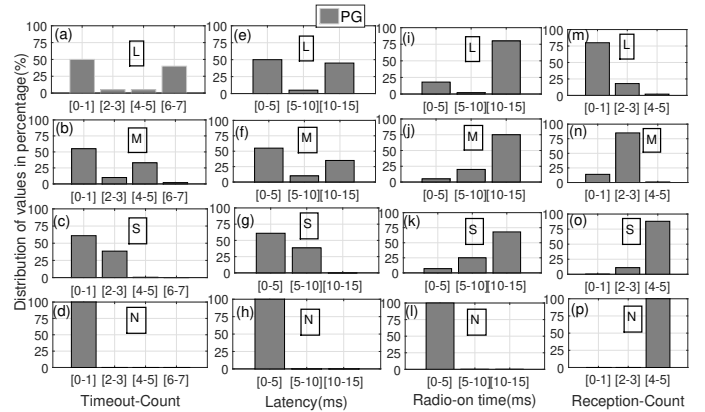}
\end{center}
\caption{Distribution of the metrics TC, LT, RO, and RXCT for vehicles of Type-L, Type-M, Type-S, and Type-N referred to as L, M, S, and N, respectively, under PG.} \label{fig:class-PG}
\end{figure}

As a first step we try to understand the impact of these types of the vehicles on the four metrics based on the already collected experiment data and the corresponding GT. Figure \ref{fig:class-PG} shows the distribution of the possible values of the four metrics (TC, LT, RO and RXCT) for four different classes. For quick comparison, we group the possible values for each of the different metrics and calculate the number of iterations of DCPhase that satisfy each of these groups for each of the four different classes of vehicles. Although the distribution of the values approximately correlate with the type of the vehicles, we found many cases where the metrics show similar behavior for different vehicle classes which naturally tends to make the automation of the classification process complex. Our detailed study and analysis of the distributions of the metrics reveal two primary sources of such ambiguity. In the following we describe them and discuss the solution approach.

\subsection{Lack of persistent communication}

DCPhase runs a normal Glossy instance under low TXP in a two-node setting. The sole intention is to make the Glossy packets get obstructed by the vehicles in a way so that the impact recorded through the features/metrics can identify the type of the vehicle. However, since there is actually no control over the exact way a Glossy packet may collide with a vehicles, it is always to good to have a sufficient number of such collision instances even for a single vehicle. To make this happen, the transmissions from the initiator node has to be persistent in nature, i.e., initiator should keep on sending the packets as soon as possible even in case it detects that the last packet was lost. 

However, Glossy being originally a multi-hop protocol, such persistent behavior is implemented only for the very first packet transmitted from the initiator. From an initiator's perspective, when it hears nothing after the its first transmission, there are two possibilities (i) first-hop nodes could not hear anything from the initiator, or, ii) the reply packet from the first-hop nodes got lost. But initiator cannot distinguish between these two cases. Therefore, it waits for a safe period called \textit{initiator-time-out} and then re-transmit the same packet. Moreover, once the first packet successfully goes through, the same mechanism cannot get repeated for the other packets. These issues substantially reduces the possible number of collision of the packets with a vehicle.

\textbf{Solution:} Dynamics of the Glossy in DCPhase in a two-node setting, is absolutely different from the same in a multi-hop setting. In DCPhase, each of the NTX transmissions can operate in a persistent mode. To enhance the scope of collision further (as explained next), we split a Glossy instance into NTX number of persistent Glossy instances each having NTX=1, with a gap between two consecutive instances. Each instance is set with a very short value of initiator-time-out.

\subsection{Interaction between DCPhase and vehicles}

\begin{figure}[htbp]
\begin{center}
\includegraphics[angle=0,width=0.5\textwidth]{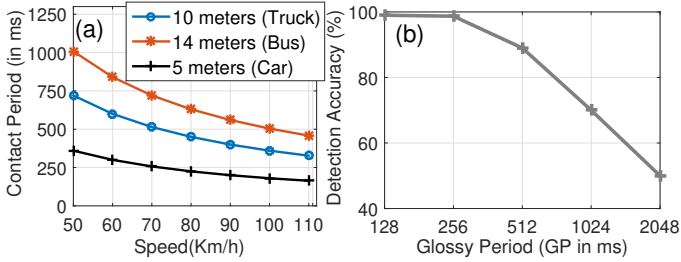}
\end{center}
\caption{Part (a) shows Contact-Period (CP) for different size and speed of the vehicles and part (b) shows the detection accuracy for different Glossy-Periods (GP) of SCPhase.}
\vspace{-0.3cm}
\label{fig:contact}
\end{figure}

The time a vehicle gets for interaction with the DCPhase is a very important issue and it depends on the size as well as speed of the vehicles. It has been seen that under low-power communication, the main disturbances happen when a vehicle crosses the line-of-sight (LoS) between the sender and the receiver \cite{sensys, comsnets-mitigation}. Let us define the quantity \textit{Contact-Period} (CP) as the amount of time a vehicle takes to cross the LoS in an MP. Figure \ref{fig:contact}(a) shows approximate CP values for different types of vehicles at different speed. It can be seen that the minimum CP considering all the realistic cases is about 250 ms. This implies that GP of SCPhase must be at least 250 ms. Since the proposed detection strategy uses a single instance of Glossy in the DCPhase, the detection accuracy drastically drops with the increase in the GP of SCPhase as shown in Figure \ref{fig:contact}(b). This is because with higher GP, most of the vehicles just passes through the MP without getting a chance to collide with the Glossy packets.

\begin{figure}[htbp]
\begin{center}
\includegraphics[angle=0,width=0.5\textwidth]{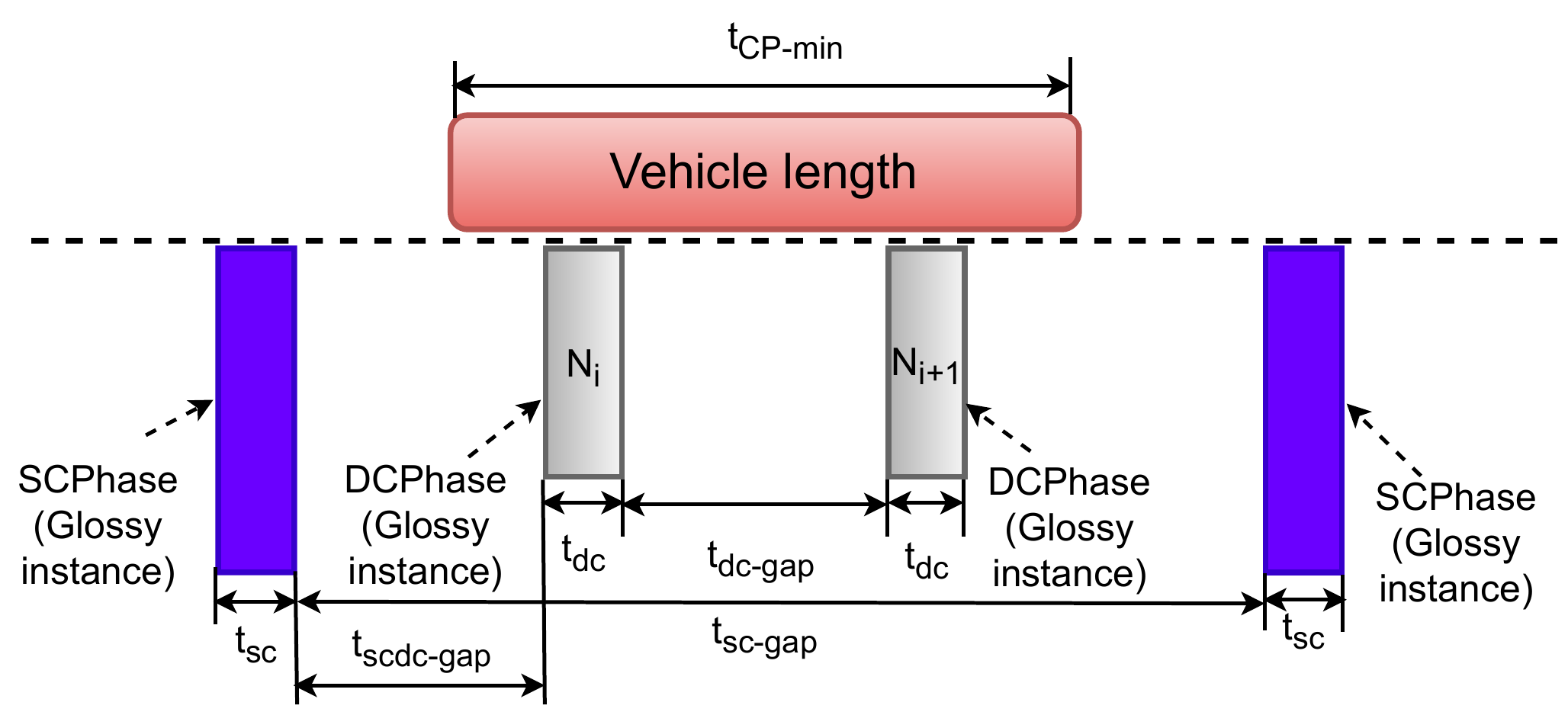}
\end{center}
\caption{Pictorial representation of the constraints summarized in Equation \ref{eq:summay-equation}.}
\vspace{-0.3cm}
\label{fig:contact-equation-derivation}
\end{figure}

\textbf{Solution:} Splitting of a single Glossy instance into multiple instances, is a step towards solving the aforementioned problem. 
The individual instances of Glossy should be scheduled in a way so that the vehicles having the minimum size as well as maximum speed (i.e., the minimum possible \textit{Contact-Period} (CP)) also should get at least one scope of interaction with the Glossy packets. In the following we derive a set of constraints that can be used as a guideline for scheduling the Glossy instances inside an SCPhase. 

Let us consider $t_{\text{CP-min}}$ to be the minimum CP. As shown in Figure \ref{fig:contact-equation-derivation}, considering two consecutive instances of Glossy with NTX=1 ($\text{N}_1$ and $\text{N}_2$), each having duration of $t_{\text{dc}}$, the gap between them, i.e., $t_{\text{dc-gap}}$ should be set in a way so that it meets the following constraint.

\begin{equation}
    t_{\text{CP-min}} > t_{\text{dc}}  + t_{\text{dc-gap}} +  t_{\text{dc}}.
\end{equation}

Similar constraint can be applied for the gap between SCPhase and the first Glossy instance in DCPhase, i.e., $t_{\text{scdc-gap}}$ as,

\begin{equation}
    t_{\text{CP-min}} > t_{\text{sc}}  + t_{\text{scdc-gap}},
\end{equation}

where $t_{\text{sc}}$ is the duration of the Glossy in SCPhase. From Equation 1 and 2, we can conclude,

\begin{equation}
    t_{\text{CP-min}} > t_{\text{dc}}  + t_{\text{dc-gap}} +  t_{\text{dc}}\geq t_{\text{sc}}  + t_{\text{scdc-gap}}.
\end{equation}

Considering a complete period of SCPhase, all the constraints can summarized as follows -

\begin{equation}
      2 t_{\text{sc}}  + t_{\text{sc-gap}} > t_{\text{CP-min}} > 2 t_{\text{dc}}  + t_{\text{dc-gap}}\geq t_{\text{sc}}  + t_{\text{scdc-gap}}.
      \label{eq:summay-equation}
\end{equation}


\begin{figure}[htbp]
\begin{center}
\includegraphics[angle=0,width=0.5\textwidth]{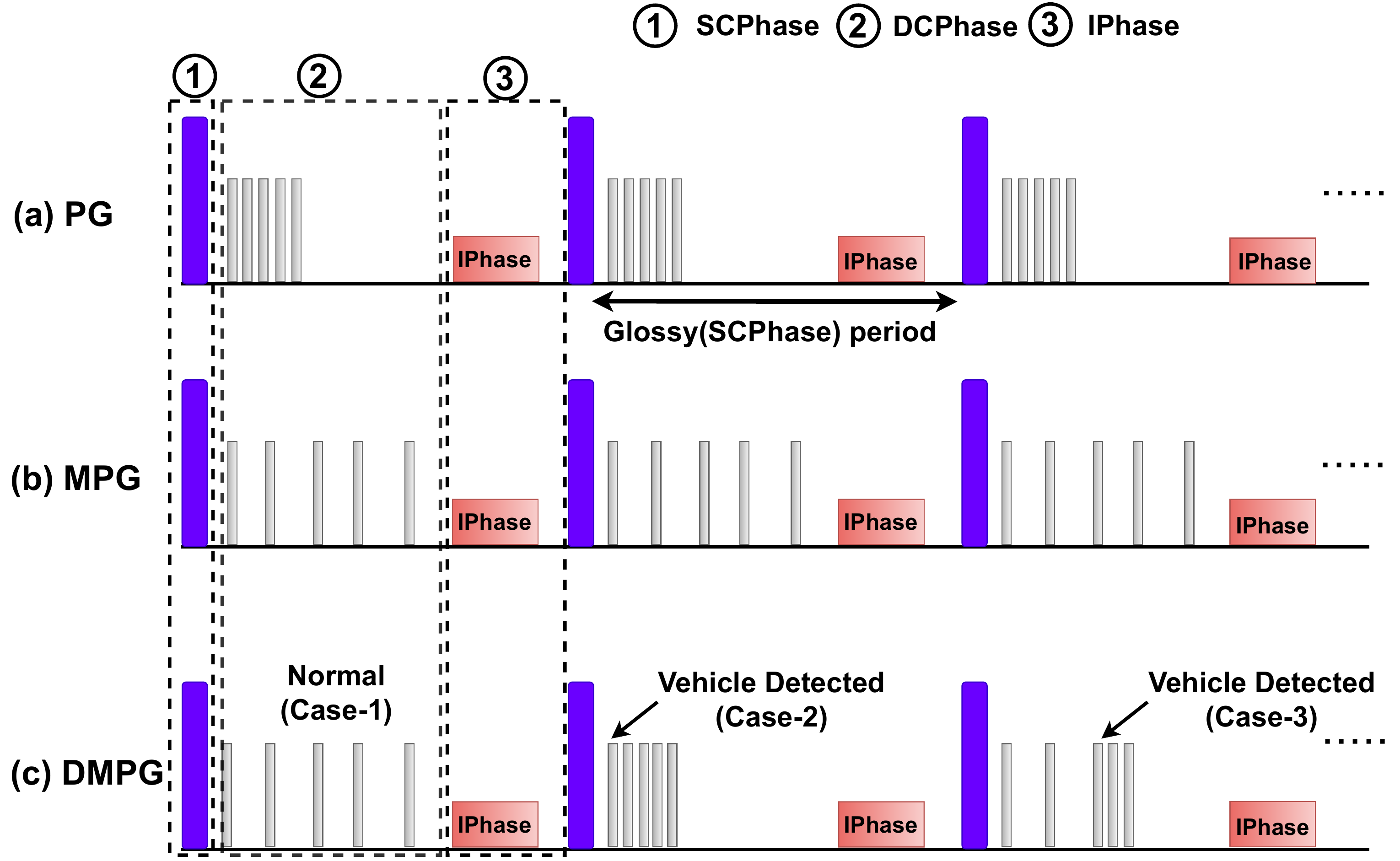}
\end{center}
\caption{Time diagram of the three strategies. (a) PG: The transmissions in DCPhase happens within a single Glossy run. (b) MPG: Instead of a single Glossy instance, a series of Glossy instances (each with NTX=1) are used to uniformly cover the SPhase. (c) DMPG: Same as MPG with more frequent execution of the Glossy instances on demand. For example, Case-1 is same as MPG, while in Case-2, the vehicle is detected in the starting and hence, all the instances are run sequentially. In Case-3, the vehicle is detected little later hence, the rest of the Glossy instances only are repeated sequentially.}
\vspace{-0.3cm}
\label{fig:design}
\end{figure}

\subsection{Design: Re-organizing DCPhase}

Based on the constraints explained above, we design two strategies as described below. 

\begin{enumerate}

\item \textit{Multiple Plain Glossy} (MPG): A consecutive NI number of instances each having NTX=1 are scheduled based on the SCPhase period. Thus, in a nutshell, the SCPhase period is equally divided into multiple parts based on $t_{\text{CP-min}}$ to balance both minimum energy consumption and enough interaction of the vehicles with Glossy instances. 

\item \textit{Dynamic Multiple Plain Glossy} (DMPG): It is supposed to begin operation as MPG. However, when a single Glossy instance collides with the vehicles and its reflected in the metrics, the other instances in the same SCPhase are also scheduled \textit{at a shorter gap} ($d_x$) instead of waiting for their original schedule as set in MPG.

\end{enumerate}

The initial version of the design where we run a single instance of Glossy without any splitting is referred hereafter as \textit{Plain Glossy} (PG). Figure \ref{fig:design} pictorially demonstrates the time-diagram of all the three strategies.


\begin{figure}[htbp]
\begin{center}
\includegraphics[width=0.5\textwidth]{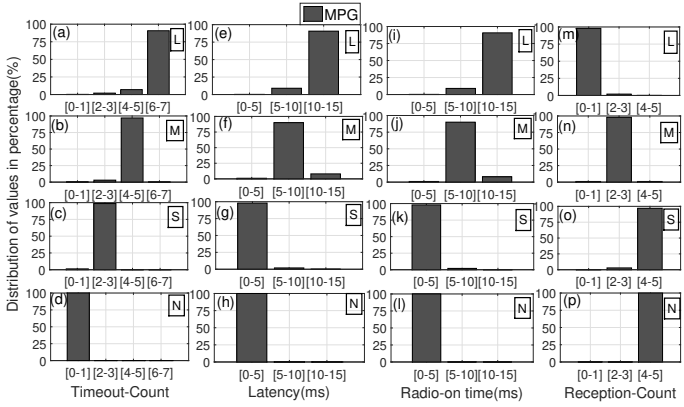}
\end{center}
\caption{Distribution of the metrics TC, LT, RO, and RXCT for vehicles of Type-L, Type-M, Type-S, and Type-N referred to as L, M, S, and N, respectively, under MPG.} \label{fig:class-MPG}
\end{figure}

\begin{figure}[htbp]
\begin{center}
\includegraphics[width=0.5\textwidth]{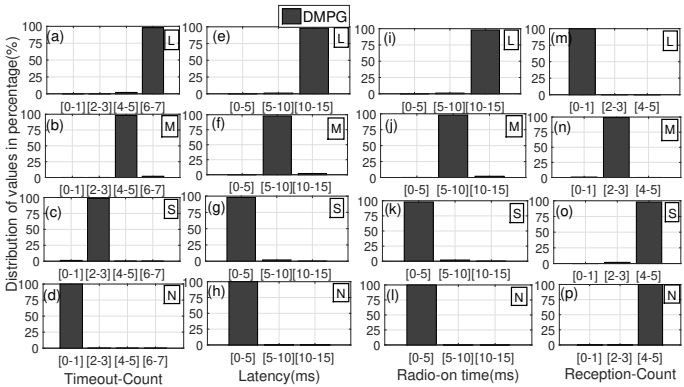}
\end{center}
\caption{Distribution of the metrics TC, LT, RO, and RXCT for vehicles of Type-L, Type-M, Type-S, and Type-N, denoted by L, M, S, and N, respectively, under DMPG.} \label{fig:class-DMPG}
\end{figure}

\subsection{Distribution of the metrics}

We repeat the outdoor experiments with MPG and DMPG. As a very first step we carry out an extensive analysis of the collected measurement data and derive the distribution of the four metrics for all the four distinct classes with the help of the GT.
Figure \ref{fig:class-MPG}(a), \ref{fig:class-MPG}(b),  \ref{fig:class-MPG}(c), and  \ref{fig:class-MPG}(d) show the distribution of TC in MPG for all the four different classes. Figure \ref{fig:class-DMPG}(a), \ref{fig:class-DMPG}(b), \ref{fig:class-DMPG}(c), and \ref{fig:class-DMPG}(d) depict the same for DMPG. The differences between the same distributions obtained from PG as shown in Figure \ref{fig:class-PG}(a), \ref{fig:class-PG}(b), \ref{fig:class-PG}(c), and \ref{fig:class-PG}(d) are very clearly visible. In particular, TC can be seen to be quite uniquely determining the type of the vehicle in DMPG. In MGP the behavior is still almost predictable while in PG it appears difficult. Similar effect can be observed in the distribution of RXCT, LT and RO as well.



\begin{table}
\begin{center}
\includegraphics[angle=0,width=0.5\textwidth]{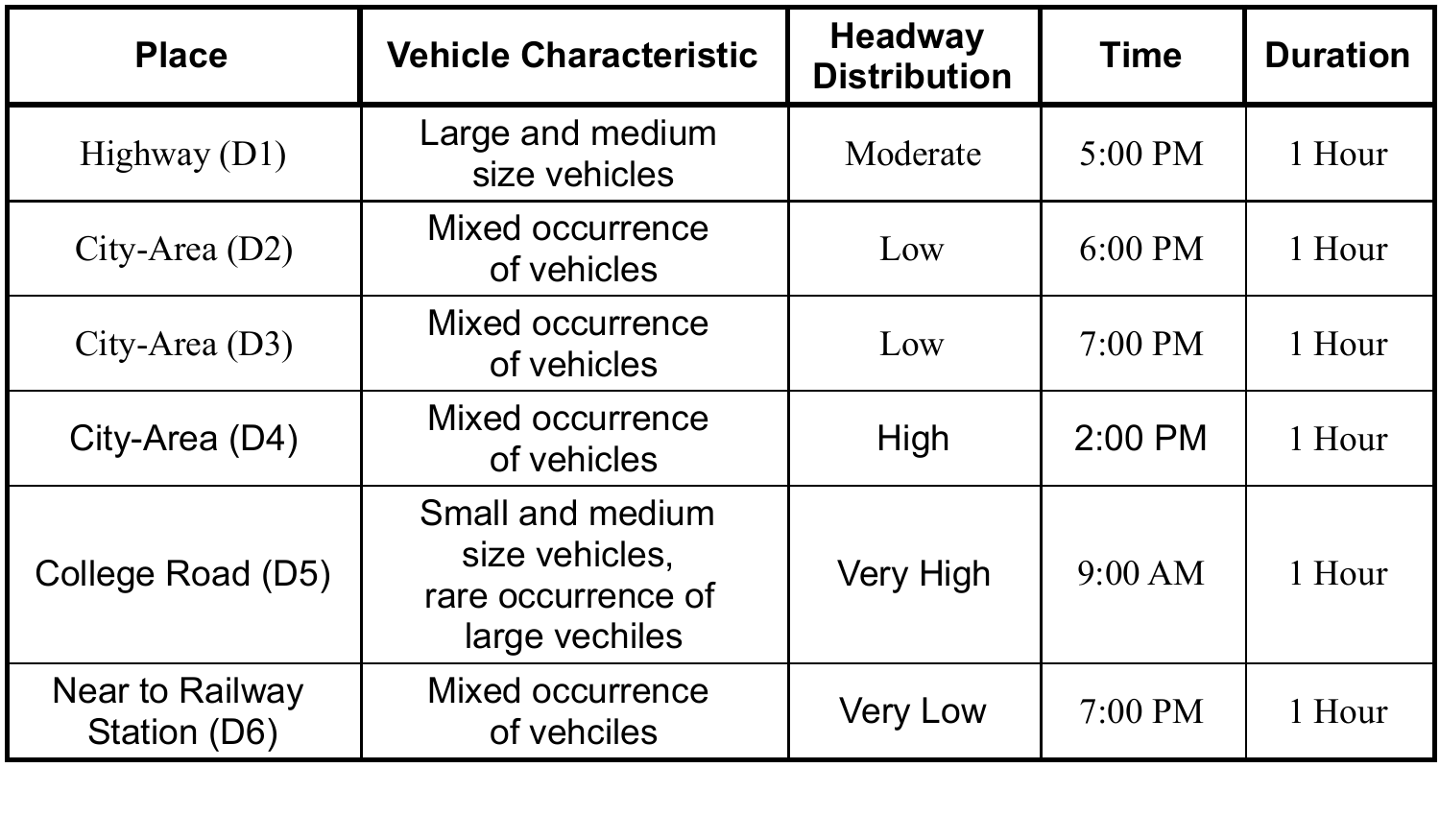}
\end{center}
\caption{Description of experimental setting, status of the traffic and vehicles during experiments.}
\label{table:Dataset}

\end{table}

\section{Evaluation} \label{sec:evaluation}

Study of the distribution of the metrics/features fundamentally provides a good confidence regarding the feasibility of classification of the type of the vehicles. Since the metrics are found to be well-distributed over different classes of the vehicles making the classification problem less complex, we use lightweight SVM as the primary means. Although some of the metrics have the capability of determining the class of a vehicle almost uniquely (as can be observed form the distribution of the metrics obtained from MPG and DMPG shown in Figure \ref{fig:class-MPG} and \ref{fig:class-DMPG}), we use a combination of all the four metrics to get higher accuracy for all possible cases throughout.

Table \ref{table:Dataset} shows the details of the measurement experiments carried out with MPG and DMPG. All the four metrics (TC, RO, RXCT and LT) are used as the set of features from each iteration of DCPhase. From GT we obtain the exact class of the vehicle for each iteration and use this information to label the data for each period of SCPhase. 

\begin{table}
\begin{center}
\includegraphics[angle=0,width=0.48\textwidth]{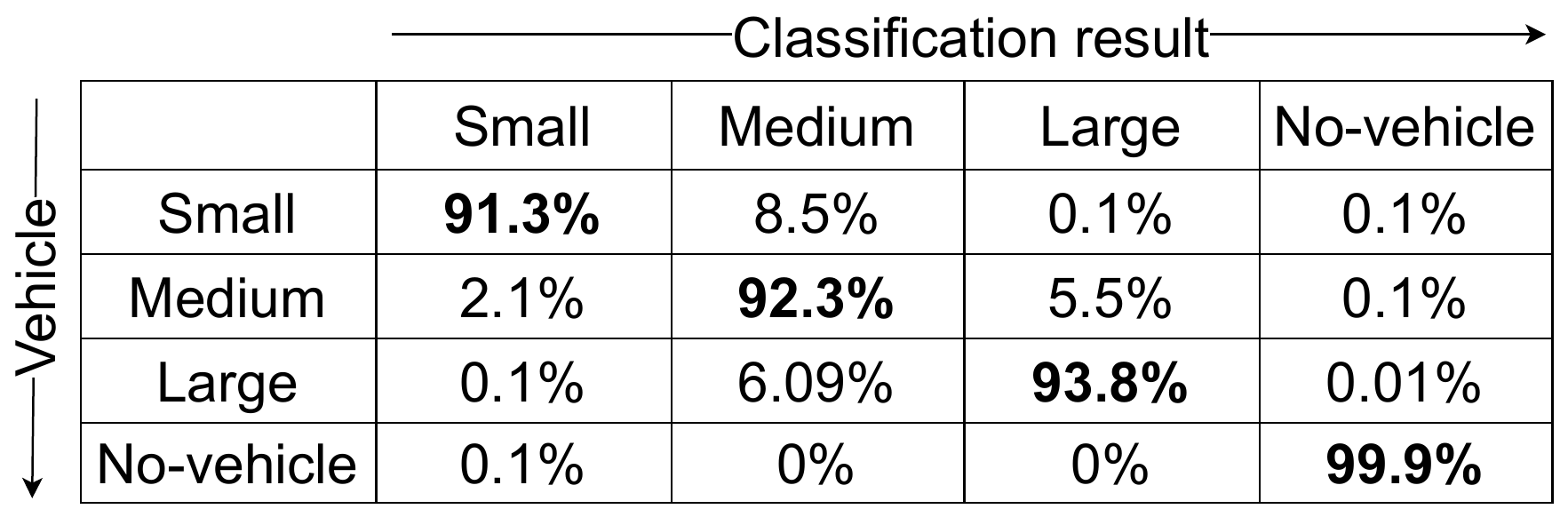}
\end{center}
\caption{Vehicle classification accuracy.}
\label{table:vehicle-classification}
\end{table}

In general, the communication link between the devices in an MP may get obstructed by multiple vehicles of different class or of same class at the same time. It depends on the traffic-intensity characterized by the \textit{headway-distribution} \cite{headway}. 
Very less such cases were found under low/moderate traffic. However, to make our analysis simpler in all our initial studies we clean the data and remove all such \textit{mixed/overlapped} cases. Finally we use the cleaned labeled data-set to train an SVM. About 80\% of the labelled data is used for training and rest is used for testing purpose where we obtain around 92.5\% mean accuracy in classifying the vehicles. Detailed results obtained from the experiments done over data-set \textit{D2} are shown in Table \ref{table:vehicle-classification}. Furthermore, to cross validate the accuracy we train the model over the union of a number of data-sets and test over a different data-set which is not used for training. Table \ref{table:crossvlidation} shows the average accuracy obtained in all these experiments.

\subsection{Performance under Mixed Setting}

Occurrence of overlapping cases were quite few under low as well as even moderate traffic. However, under high traffic (low-headway), we find a noticeable amount of such overlaps. None of the existing works in this direction deals with such complex situations. However, we do further investigation as detailed below to check the efficacy of our proposed design under such cases.


We introduce three more classes - \textit{L-mix}, \textit{M-mix} and \textit{S-mix}. L-mix represents the cases where multiple vehicles were present with \textit{at least one Type-L vehicle}. M-mix represents the cases where there is \textit{at least one Type-M or Type-S vehicles and no Type-L vehicles}. S-mix denotes the cases where \textit{only multiple Type-S vehicles are present}. The same data-sets collected earlier are now re-labeled considering total 7 classes (including Type-N). For each data-set we train an SVM with all these 7 classes. Subsequently, it is tested based on the cleaned version of some other data set with four classes. 

Table \ref{table:highspeed} summarizes the average accuracy values when training is done on 7 classes over complete data-set D2 while testing is done on the cleaned version of data-set D4 over each of the four basic classes. We observe that, while detection accuracy stays unaltered, a good many cases of Type-S are wrongly classified as S-mix. Similarly, some occurrence of Type-M are classified as M-mix. We also observe that two overlapping occurrences of Type-M are sometime classified as Type-L as well as two overlapping occurrences of Type-S vehicles are classified as Type-M, as we discussed above. 

\begin{figure}[htbp]
\begin{center}
\includegraphics[width=0.35\textwidth]{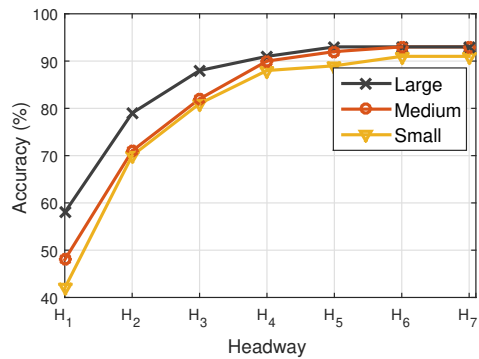}
\end{center}
\caption{Shows the average classification accuracy under different traffic rate (represented by headway, in sec), where the labels in X-axis are as follows $H_{1}=0.25$, $H_{2}=0.5$, $H_{3}=0.75$, $H_{4}=1$, $H_{5}=2$, $H_{6}=3$, $H_{7}=4$ (all are in sec).}
\vspace{-0.3cm}
\label{fig:accuracy-headway}
\end{figure}

The mean accuracy of the overall classification, thus, degrades due to overlapping occurrences of the vehicles. For a complete understanding we derive the headway-distribution and its average from different segments of our collected data with the help of GT. Figure \ref{fig:accuracy-headway} shows the accuracy obtained in vehicle classification for all these different average headway. It can be seen that classification accuracy gracefully degrades with  lower mean-headway, (i.e., under higher traffic intensity).



\begin{table}
\begin{center}
\includegraphics[angle=0,width=0.4\textwidth]{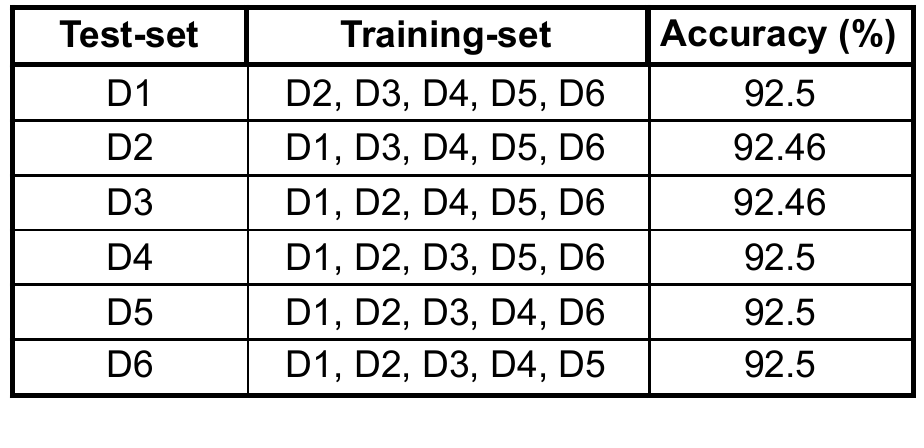}
\end{center}
\caption{Cross-validation of average classification accuracy in the experiment described in Section \ref{sec:vehicle} over few different pairs of data-sets.}
\vspace{-0.5cm}
\label{table:crossvlidation}

\end{table}

\begin{table}
\begin{center}
\includegraphics[angle=0,width=0.5\textwidth]{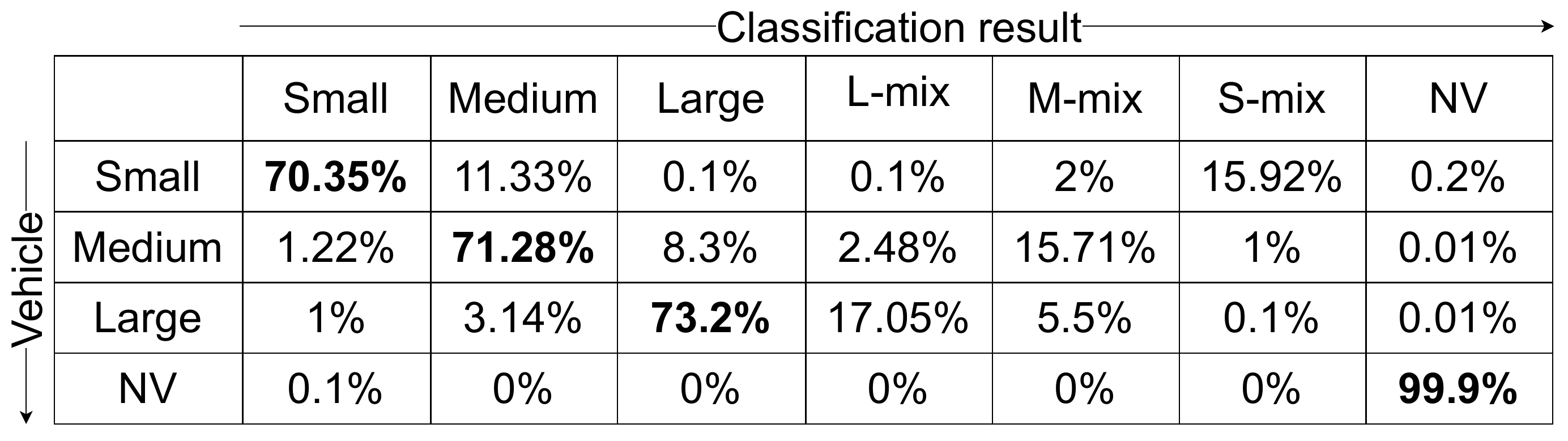}
\end{center}
\caption{Detailed classification accuracy when training is done using D2 (seven classes) and testing is done over D4 (four classes).}
\label{table:highspeed}
\end{table}

\section{Monitoring Wide area}\label{sec:widearea}

\begin{figure}[htbp]
\begin{center}
\includegraphics[width=0.5\textwidth]{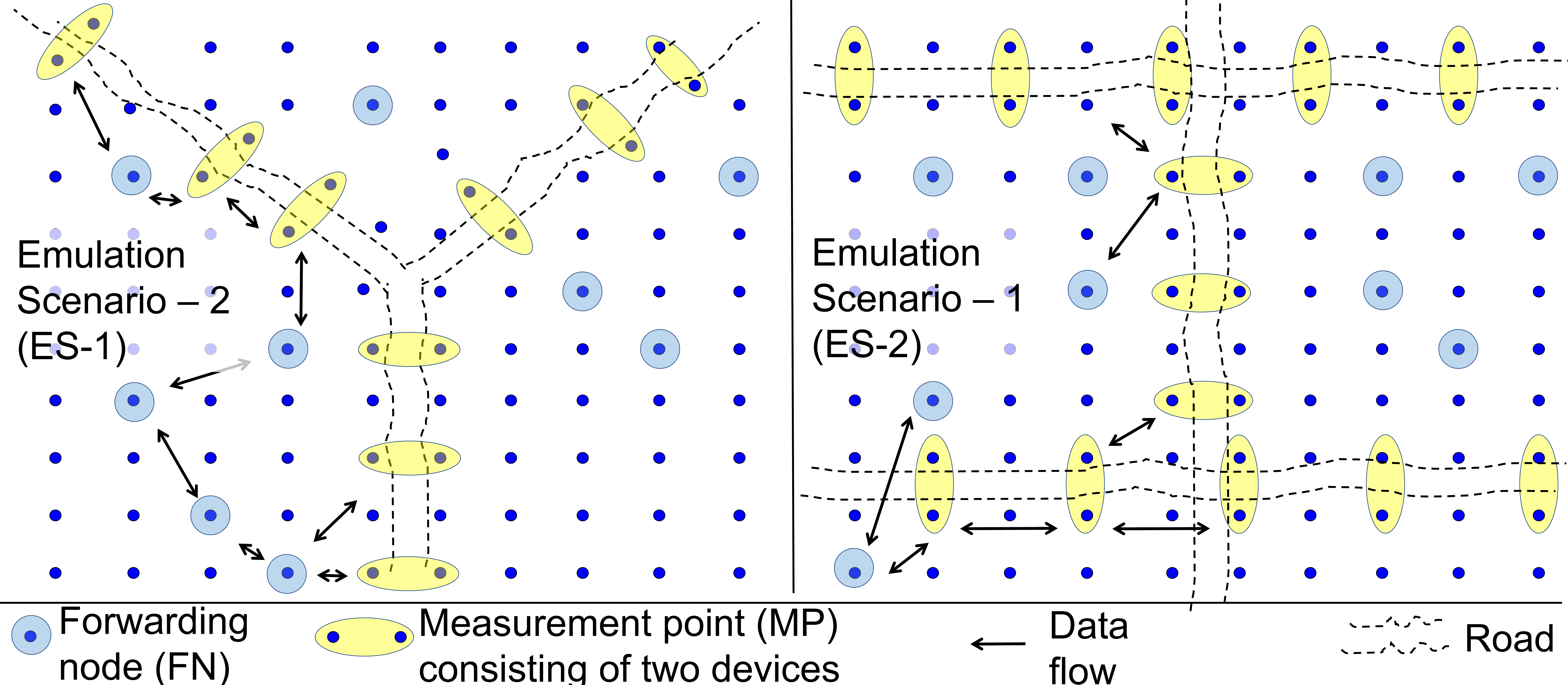}
\end{center}
\caption{Emulation scenarios.} \label{fig:sim}
\end{figure}

In this section we provide a study of the proposed strategy regarding its low-power and time-correlated operation over wide-area. We assume that the roads are covered by multiple MPs installed at an equal and appropriate distance from each other. The IoT-devices in MP and FN together form a low-power \textit{ad hoc} network. We use the \textit{Contiki network emulator Cooja} \cite{LCN-cooja} for all these experiments. Two emulation scenarios, referred by ES-1 and ES-2 (shown in Figure \ref{fig:sim}), are created mimicking two different types of physical arrangements of the roads with total 100 TelosB devices forming a 2D-grid like structure over an area of $2000 X 2000$ sq. meter and $1000 X 1000$ sq. meter respectively. Inter-MP distance is set as 50 meter and 75 meter, respectively in ES-1 and ES-2. 

The metrics RXCT, RO, and LT (see Section \ref{sec:metric}) are measured in both DCPhase and SCPhase. Instead of TC, we measure a new metric \textit{Hop-Count} as described below.

\textbf{Hop-Count}: In an instance of Glossy, the parameter \textit{relay-count} is used by every node to keep track of the number of times a packet has been transmitted/forwarded \cite{glossy}. The value of the relay-count obtained from the very first packet-reception in a node indicates a rough hop-position of the node in the network w.r.t. the initiator. We use this value as the \textit{Hop-Count} (HC).

Each of these 4 metrics are measured in each IoT-device over approx. 1000 iterations of the experiments. The results show the average and the standard deviations (in the form of error-bars).

\begin{figure}[htbp]
\begin{center}
\includegraphics[width=0.5\textwidth]{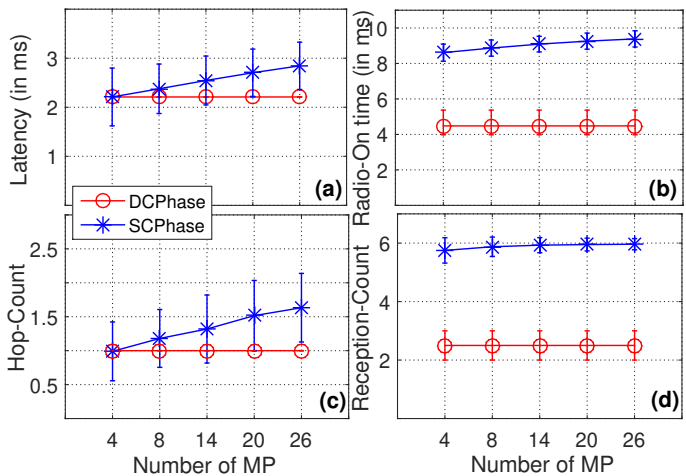}
\end{center}\caption{Effect of the number of MP on the performance of the system. Part (a), (b), (c), (d) show the plot of LT, RO, HC and RXCT, in ES-1 for different number of MPs.}
\label{fig:effectofMPs}
\end{figure}

\textbf{Effect of number of MPs}: First we experiment with a skeleton network created solely by the MPs (i.e., no FNs). Figure \ref{fig:effectofMPs} shows the plots of HC, RXCT, LT and RO for both SCPhase and DCPhase (in PG) for varying number of MPs under ES-1. The higher the no of MPs, the higher the diameter of the network which is reflected with almost linear rise in RO, LT, and HC. RXCT remains the same implying a smooth functioning of the system. Note that the change in the number of MPs affects only the results in SCPhase, not in DCPhase.




\begin{figure}[htbp]
\begin{center}
\includegraphics[width=0.5\textwidth]{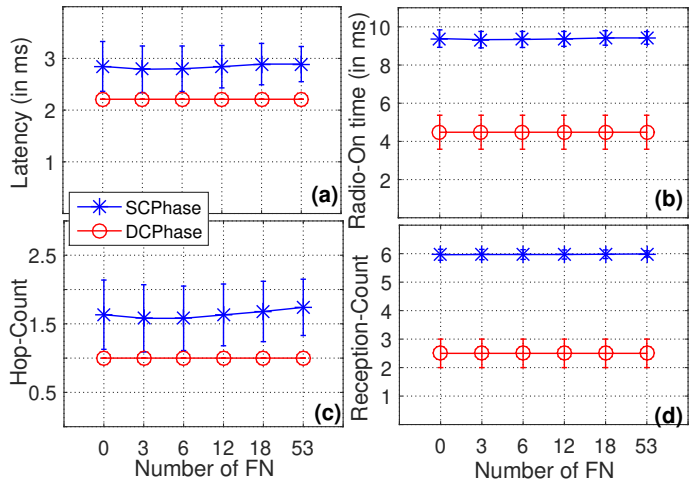}
\end{center}
\caption{Effect of different number of FNs in ES-1.} \label{fig:effectofFN}
\end{figure}

\textbf{Effect of FN:} We experiment with a set of 20 number of MPs and gradually increase the number of FNs in various ways. Figure \ref{fig:effectofFN} shows the average HC, RO, RXCT and LT in each of these cases in both SCPhase and DCPhase in ES-1. It can be observed that there is almost no effect on the system performance as long as the MPs are installed maintaining proper inter-MP distance so that the network remains connected. 
Due to its inherent robustness, Glossy is thus quite able to sustain even with a skeletal setting of the MPs and minimum number of FN. Lower number of FN also implies lesser cost of deployment too. 

\begin{figure}[htbp]
\begin{center}
\includegraphics[width=0.5\textwidth]{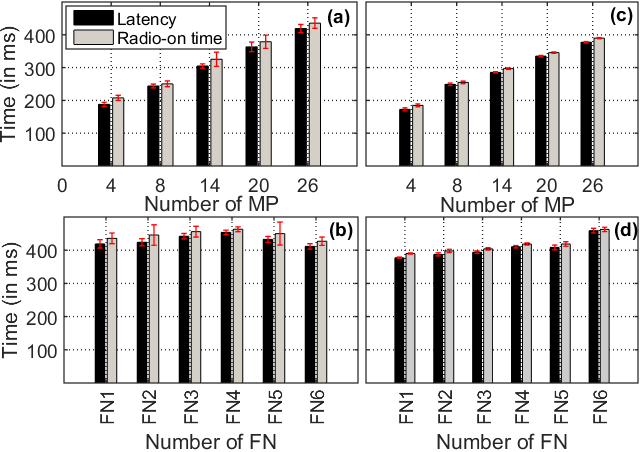}
\end{center}
\caption{Effect of different number of MPs and FNs over the performance over IPhase. FN1: 0, FN2: 3 (L), FN3: 12 (L), FN4: 6 (L, R), FN5: 12 (L, R), FN: 48. L and R indicate the number FNs are taken from left side and right side of the main road in emulation setting.
}
\vspace{-0.6cm}
\label{fig:datasharing}
\end{figure}

\textbf{IPhase:} Finally, we also experiment with a possible realization of IPhase. In particular, we integrate the protocol Chaos \cite{chaos} for all-to-all data-sharing in the IPhase. Figure \ref{fig:datasharing}(a), \ref{fig:datasharing}(b) and \ref{fig:datasharing}(c), \ref{fig:datasharing}(d) shows the LT and RO in IPhase for varying number of MP and FNs under ES-1 and ES-2, respectively. It can be observed that RO and LT in IPhase almost linearly rise with the number of MPs in both the settings while does not show any significant effect with the rise in the number of FNs. Note that only the MPs act as the source of data while FNs only act as forwarding devices.

All the above mentioned experiments were also repeated under ES-2 and similar trends in the metrics were observed.

\section{Conclusion}\label{sec:conclusion}

In this work we introduce a novel, low-cost, flexible and easy-to-install mechanism for real-time detection and classification of the vehicles over wide-area. It does not need any specialized infrastructure as well as use only off-the-shelf devices to solve a vital problem in smart-traffic management. Through extensive outdoor measurement studies using real IoT-devices we show the effectiveness of the principles behind the proposed strategy. 

Rigorous evaluation of the proposed strategy based on outdoor experiments and empirical data collected from various urban places demonstrates that it can detect and classify with quite high accuracy (above 93\%) under low and moderate traffic. While behavior of any existing  strategy under high traffic is rarely addressed in the literature, we show that our strategy degrades gracefully with increase in the traffic intensity because of overlapping occurrences of the vehicles. The performance of the strategy can improve with appropriate incorporation of low-cost multimedia sensors, and sophisticated lightweight edge-computing tools. Moreover, algorithmic paradigm such as divide, conquer and merge can be applied to extend the scope of the proposed framework to cover even larger area. We consider these issues as part of our immediate future steps in this direction.

\bibliographystyle{abbrv}
\bibliography{vehicle}

\begin{thebibliography}{10}

\bibitem{Bajwa-pavement}
R.~Bajwa, R.~Rajagopal, P.~Varaiya, and R.~Kavaler.
\newblock In-pavement wireless sensor network for vehicle classification.
\newblock {\em Proceedings of the 10th ACM/IEEE International Conference on
  Information Processing in Sensor Networks}, pages 85--96, 2011.

\bibitem{Bluetooth}
M.~Bernas, B.~P{\l}aczek, and W.~Korski.
\newblock Wireless network with bluetooth low energy beacons for vehicle
  detection and classification.
\newblock In P.~Gaj, M.~Sawicki, G.~Suchacka, and A.~Kwiecie{\'{n}}, editors,
  {\em Computer Networks}, pages 429--444, Cham, 2018. Springer International
  Publishing.

\bibitem{its-cloud}
S.~Bitam and A.~Mellouk.
\newblock Its-cloud: Cloud computing for intelligent transportation system.
\newblock In {\em 2012 IEEE Global Communications Conference (GLOBECOM)}, pages
  2054--2059, 2012.

\bibitem{camera}
Z.~{Chen}, T.~{Ellis}, and S.~A. {Velastin}.
\newblock Vehicle detection, tracking and classification in urban traffic.
\newblock In {\em 2012 15th International IEEE Conference on Intelligent
  Transportation Systems}, pages 951--956, 2012.

\bibitem{headway}
R.~J. Cowan.
\newblock Useful headway models.
\newblock {\em Transportation Research}, 9(6):371--375, 1975.

\bibitem{camera-CNN}
Z.~Dong, Y.~Wu, M.~Pei, and Y.~Jia.
\newblock Vehicle type classification using a semisupervised convolutional
  neural network.
\newblock {\em IEEE Transactions on Intelligent Transportation Systems},
  16(4):2247--2256, 2015.

\bibitem{LWB}
F.~Ferrari, M.~Zimmerling, L.~Mottola, and L.~Thiele.
\newblock Low-power wireless bus.
\newblock In {\em Proceedings of the 10th ACM Conference on Embedded Network
  Sensor Systems}, SenSys '12, page 1–14, New York, NY, USA, 2012.
  Association for Computing Machinery.

\bibitem{glossy}
F.~Ferrari, M.~Zimmerling, L.~Thiele, and O.~Saukh.
\newblock Efficient network flooding and time synchronization with glossy.
\newblock In {\em Proceedings of ACM/IEEE IPSN}, pages 73--84, 2011.

\bibitem{application}
S.~Garg, P.~Singh, P.~Ramanathan, and R.~Sen.
\newblock Vividhavahana: Smartphone based vehicle classification and its
  applications in developing region.
\newblock In {\em Proceedings of the 11th International Conference on Mobile
  and Ubiquitous Systems: Computing, Networking and Services}, MOBIQUITOUS '14,
  page 364–373, Brussels, BEL, 2014. ICST (Institute for Computer Sciences,
  Social-Informatics and Telecommunications Engineering).

\bibitem{CTP}
O.~Gnawali, R.~Fonseca, K.~Jamieson, D.~Moss, and P.~Levis.
\newblock Collection tree protocol.
\newblock In {\em Proceedings of the 7th ACM Conference on Embedded Networked
  Sensor Systems}, SenSys '09, page 1–14, New York, NY, USA, 2009.
  Association for Computing Machinery.

\bibitem{RSSI-BASED-VTC}
M.~{Haferkamp}, M.~{Al-Askary}, D.~{Dorn}, B.~{Sliwa}, L.~{Habel},
  M.~{Schreckenberg}, and C.~{Wietfeld}.
\newblock Radio-based traffic flow detection and vehicle classification for
  future smart cities.
\newblock In {\em 2017 IEEE 85th Vehicular Technology Conference (VTC Spring)},
  pages 1--5, 2017.

\bibitem{loop-detector}
S.~{Jeng} and L.~{Chu}.
\newblock A high-definition traffic performance monitoring system with the
  inductive loop detector signature technology.
\newblock In {\em 17th International IEEE Conference on Intelligent
  Transportation Systems (ITSC)}, pages 1820--1825, 2014.

\bibitem{chaos}
O.~Landsiedel, F.~Ferrari, and M.~Zimmerling.
\newblock Chaos: Versatile and efficient all-to-all data sharing and in-network
  processing at scale.
\newblock In {\em Proceedings of ACM SenSys}, page~1, 2013.

\bibitem{capture-effect}
K.~Leentvaar and J.~Flint.
\newblock The capture effect in fm receivers.
\newblock {\em IEEE Transactions on Communications}, 24(5):531--539, 1976.

\bibitem{RSSI-based}
X.~{Li} and J.~{Wu}.
\newblock A new method and verification of vehicles detection based on rssi
  variation.
\newblock In {\em 2016 10th International Conference on Sensing Technology
  (ICST)}, pages 1--6, 2016.

\bibitem{camera-couting}
M.~Liang, X.~Huang, C.-H. Chen, X.~Chen, and A.~Tokuta.
\newblock Counting and classification of highway vehicles by regression
  analysis.
\newblock {\em IEEE Transactions on Intelligent Transportation Systems},
  16(5):2878--2888, 2015.

\bibitem{survey-sensors}
T.~Lin, H.~Rivano, and F.~Le~Mouël.
\newblock A survey of smart parking solutions.
\newblock {\em IEEE Transactions on Intelligent Transportation Systems},
  18(12):3229--3253, 2017.

\bibitem{accelerometer-ITS}
W.~Ma, D.~Xing, A.~McKee, R.~Bajwa, C.~Flores, B.~Fuller, and P.~Varaiya.
\newblock A wireless accelerometer-based automatic vehicle classification
  prototype system.
\newblock {\em IEEE Transactions on Intelligent Transportation Systems},
  15(1):104--111, 2014.

\bibitem{power}
D.~Minoli, K.~Sohraby, and B.~Occhiogrosso.
\newblock Iot considerations, requirements, and architectures for smart
  buildings—energy optimization and next-generation building management
  systems.
\newblock {\em IEEE Internet of Things Journal}, 4(1):269--283, 2017.

\bibitem{infrared}
E.~{Odat}, J.~S. {Shamma}, and C.~{Claudel}.
\newblock Vehicle classification and speed estimation using combined passive
  infrared/ultrasonic sensors.
\newblock {\em IEEE Transactions on Intelligent Transportation Systems},
  19(5):1593--1606, 2018.

\bibitem{LCN-cooja}
F.~Osterlind, A.~Dunkels, J.~Eriksson, N.~Finne, and T.~Voigt.
\newblock Cross-level sensor network simulation with cooja.
\newblock In {\em Proceedings. 2006 31st IEEE Conference on Local Computer
  Networks}, pages 641--648, 2006.

\bibitem{Piezo-Electric-Sensor}
S.~Rajab, A.~Mayeli, and H.~Refai.
\newblock Vehicle classification and accurate speed calculation using
  multi-element piezoelectric sensor.
\newblock pages 894--899, 06 2014.

\bibitem{vibration-2}
J.~Rivas, R.~Wunderlich, and S.~J. Heinen.
\newblock Road vibrations as a source to detect the presence and speed of
  vehicles.
\newblock {\em IEEE Sensors Journal}, 17(2):377--385, 2017.

\bibitem{comsnets-lqi-prr}
S.~{Roy}, R.~{Sen}, S.~{Kulkarni}, P.~{Kulkarni}, B.~{Raman}, and L.~K.
  {Singh}.
\newblock Wireless across road: Rf based road traffic congestion detection.
\newblock In {\em 2011 Third International Conference on Communication Systems
  and Networks (COMSNETS 2011)}, pages 1--6, 2011.

\bibitem{sensys}
R.~Sen, A.~Maurya, B.~Raman, R.~Mehta, R.~Kalyanaraman, N.~Vankadhara, S.~Roy,
  and P.~Sharma.
\newblock Kyun queue: A sensor network system to monitor road traffic queues.
\newblock In {\em Proceedings of the 10th ACM Conference on Embedded Network
  Sensor Systems}, SenSys '12, page 127–140, New York, NY, USA, 2012.
  Association for Computing Machinery.

\bibitem{horn-ok-please}
R.~Sen, B.~Raman, and P.~Sharma.
\newblock Horn-ok-please.
\newblock In {\em Proceedings of the 8th International Conference on Mobile
  Systems, Applications, and Services}, MobiSys '10, page 137–150, New York,
  NY, USA, 2010. Association for Computing Machinery.

\bibitem{volume-measurement}
C.~Shekhar and S.~Saha.
\newblock Iot-assisted low-cost traffic volume measurement and control.
\newblock In {\em 2022 14th International Conference on COMmunication Systems
  NETworkS (COMSNETS)}, pages 806--811, 2022.

\bibitem{comsnets-mitigation}
C.~Shekhar, S.~Saha, and M.~C. Chan.
\newblock Mitigating adversities in urban iot-setup: A sensor assisted
  approach.
\newblock In {\em 2021 International Conference on COMmunication Systems
  NETworkS (COMSNETS)}, pages 413--420, 2021.

\bibitem{vibration}
M.~{Stocker}, M.~{Rönkkö}, and M.~{Kolehmainen}.
\newblock Situational knowledge representation for traffic observed by a
  pavement vibration sensor network.
\newblock {\em IEEE Transactions on Intelligent Transportation Systems},
  15(4):1441--1450, 2014.

\bibitem{optical}
Z.~{Sun}, G.~{Bebis}, and R.~{Miller}.
\newblock On-road vehicle detection using optical sensors: a review.
\newblock In {\em Proceedings. The 7th International IEEE Conference on
  Intelligent Transportation Systems (IEEE Cat. No.04TH8749)}, pages 585--590,
  2004.

\bibitem{magnet-2}
H.~Tafish, W.~Balid, and H.~H. Refai.
\newblock Cost effective vehicle classification using a single wireless
  magnetometer.
\newblock In {\em 2016 International Wireless Communications and Mobile
  Computing Conference (IWCMC)}, pages 194--199, 2016.

\bibitem{magnetometer}
S.~{Taghvaeeyan} and R.~{Rajamani}.
\newblock Portable roadside sensors for vehicle counting, classification, and
  speed measurement.
\newblock {\em IEEE Transactions on Intelligent Transportation Systems},
  15(1):73--83, 2014.

\bibitem{VTC-2020}
H.~B. {Tulay} and C.~E. {Koksal}.
\newblock Increasing situational awareness in vehicular networks: Passive
  traffic sensing based on machine learning.
\newblock In {\em 2020 IEEE 91st Vehicular Technology Conference
  (VTC2020-Spring)}, pages 1--7, 2020.

\bibitem{VTC-MASS}
M.~{Won}, S.~{Sahu}, and K.~{Park}.
\newblock Deepwitraffic: Low cost wifi-based traffic monitoring system using
  deep learning.
\newblock In {\em 2019 IEEE 16th International Conference on Mobile Ad Hoc and
  Sensor Systems (MASS)}, pages 476--484, 2019.

\bibitem{witraffic}
M.~{Won}, S.~{Zhang}, and S.~H. {Son}.
\newblock Witraffic: Low-cost and non-intrusive traffic monitoring system using
  wifi.
\newblock In {\em 2017 26th International Conference on Computer Communication
  and Networks (ICCCN)}, pages 1--9, 2017.

\bibitem{cloud-load}
L.~Zhao, J.~Wang, J.~Liu, and N.~Kato.
\newblock Optimal edge resource allocation in iot-based smart cities.
\newblock {\em IEEE Network}, 33(2):30--35, 2019.

\end{thebibliography}
\end{document}